\documentclass[aps,prx,twocolumn,showpacs]{revtex4-1}
\usepackage{bm}
\usepackage{mathrsfs}
\usepackage{amsmath}
\usepackage{amssymb}
\usepackage{graphicx}
\usepackage{amsfonts}
\usepackage{amsthm}
\usepackage{color}
\usepackage{dcolumn}
\usepackage{txfonts}
\usepackage[caption=false]{subfig}
\DeclareMathOperator{\Ai}{Ai}
\DeclareMathOperator{\Bi}{Bi}

\begin{document}

\title{Analytical Approximations for Generalized Landau-Zener Transitions in Multi-level Non-Hermitian Systems}
\author{Chon-Fai Kam}
\email{Email: dubussygauss@gmail.com}
\affiliation{Department of Mathematics, Faculty of Science and Technology,
University of Macau, Avenida da Universidade, Taipa, Macau, China}
\author{Yang Chen}
\email{Email: yangbrookchen@yahoo.co.uk}
\affiliation{Department of Mathematics, Faculty of Science and Technology,
University of Macau, Avenida da Universidade, Taipa, Macau, China}

\begin{abstract}
We study the dynamics of non-adiabatic transitions in non-Hermitian multi-level parabolic models where the separations of the diabatic energies are quadratic function of time. The model Hamiltonian has been used to describe the non-Hermitian dynamics of two pairs of coupled cavities. In the absence of the coupling between any two pairs of cavities, the wave amplitudes within each subsystem are described by the tri-confluent Heun functions. When all the couplings between the cavities are present, we reduce the dynamics into a set of two coupled tri-confluent Heun equations, from which we derive analytical approximations for the wave amplitudes at different physical limits.
\end{abstract}

\maketitle

\section{Introduction}
Since the dawn of the twentieth century, quantum mechanics has been the foundation of modern technology from the electronic computers to the most precise atomic clock. One of the basic principles of quantum mechanics is that the spectra of atoms are real and the time evolution of wave functions is unitary, and thus the total probability is conserved. As such, it was once widely believed that the Hamiltonian which describes the time evolution of any physical system has to be self-adjoint or Hermitian \cite{shankar2012principles}. However, over the years, people started to realize that the Hermitian law is not unbreakable, as the deviation of it does not directly implies complex-valued spectra energies. Since Bender and Boettcher's groundbreaking work \cite{bender1998real, bender1999pt}, a new principle started to be affirmed is that real energy spectra of a physical system are not ensured by the Hermitian property, but rather ensured by the partity-time ($\mathcal{P}\mathcal{T}$) symmetry. The fundamental difference between Hermitian system and non-Hermitian $\mathcal{P}\mathcal{T}$-symmetric system is that the former one can only be used to describes closed systems that have no exchange with the outer environment, but the later one can be used to describe open systems with two coupled subsystems, each of which is in contact with the outer environment, but the probability in one subsystem with gain compensates another subsystem with loss, so that the entire system is in a dynamical equilibrium \cite{moiseyev2011non, bender2018pt}. Over the decades, the new principle of pseudo-Hermiticity under $\mathcal{P}\mathcal{T}$-symmetry has opened up countless new opportunities, and has also revealed various application in modern technology which ranges from optics \cite{ruter2010observation} to 

One of the most interesting effects of non-Hermitian systems is that the $\mathcal{P}\mathcal{T}$-symmetry leads to a new type of spectral degeneracies, known as the exceptional points, at which not only a finite numbers of eigenvalues coincide, but the associated eigenstates also coincide \cite{bender2018pt}. In contrast to the spectral degeneracies of Hermitian systems, at which the eigenstates can be chosen to be orthogonal to each other, the spectral degeneracies in $\mathcal{P}\mathcal{T}$-symmetric systems are peculiar as certain eigenstates are completely parallel and the Hamiltonian matrix becomes defective at the exceptional points \cite{el2018non}. This intriguing properties of non-Hermitian physics give rise to many counterintuitive features. For example, a general $\mathcal{P}\mathcal{T}$-symmetric Hamiltonian may undergo a spontaneous symmetry breaking phase transition, beyond which complex eigenvalues emerge \cite{bender2003must, ozdemir2019parity}. In particular, when encircling an exceptional point, an unconventional level-crossing behavior will appear, along with a phase change of one eigenstate but not of the other \cite{heiss1990avoided, heiss2000repulsion}.

In ordinary Hermitian quantum mechanics, there is a fundamental physical process called the Landau-Zener transition, which describes the transition between two energy levels of a quantum system directly driven through an avoided crossing \cite{landau1932theoryI, landau1932theory, stuckelberg1932theory}. The Landau-Zener transition assumes a constant coupling between bare states in the diabatic basis and a linearly varying separation of diabatic energies \cite{wittig2005landau}, which can be exactly solved by the parabolic cylinder functions \cite{zener1932non} or the integral representation method \cite{majorana1932atomi}. Although the model of Landau-Zener transition has achieved great success over the years, there are indeed cases where the assumption of linear crossing between the diabatic states becomes no longer valid. For the cases in which the crossing points merge together as a result of external fields, the Landau-Zener linearization fails, and the linear-dependence of diabatic energies has to be replaced by a parabolic or superparabolic one \cite{garraway1995wave}. Interestingly, the tunneling dynamics for the parabolic and cubic models can still be exactly solved by the tri-confluent and bi-confluent Heun functions \cite{heun1888theorie, ronveaux1995heun}. One can still express the tunneling probability via the Stokes constants by using the Zhu-Nakamura formula \cite{zhu1992twoII, zhu1993two}.

Compared to the two-state scenario, the research of generalized Landau-Zener transition for complicated systems with more than two states, even in ordinary Hermitian quantum mechanics, has been arduous and in most cases inconclusive. The main reason is that in conventional Landau-Zener transition, the coupling equations which govern the non-adiabatic transition amplitudes between the two energy levels can be reduced to a single second-order differential equation, e.g., the parabolic cylinder function for linear separation of diabatic energies \cite{zener1932non}, or the confluent Heun functions for quadratic and cubic separations of diabatic energies \cite{kam2020analytical, kam2021analytical}. In contrast, in the general multi-state scenario, if one attempts to reduce the coupling equations which govern the non-adiabatic transition amplitudes between different energy levels in a single equation, one would probably obtain an ordinary differential equation with order greater than two. Compared to those of conventional second order differential equations, the analytic properties of solutions of differential equations with order greater than two are harder to obtain by regular methods like asymptotic analysis. The essential difficulty lies in the fact that the Stokes curves for conventional second order differential equations are straight lines which never cross each other, but those for differential equations with order greater than two are non-straight lines which always cross each other unavoidably. This simple fact results in the breakdown of the standard connection formula near the crossing points of the Stokes curve. In this regard, the asymptotic WKB solutions of generalized Landau-Zener non-adiabatic transitions for multi-state systems are in general hard to obtain, if not totally impossible.

Despite its evident importance, the non-Hermitian generalization of the two-level Landau-Zener transition has only recently been analyzed by Longstaff \cite{longstaff2019nonadiabatic}, the associated non-Hermitian Landau-Zener-Stuckelberg interferometry was analyzed by Shen \cite{shen2019landau}, and the non-Hermitian generalization of the parabolic and super-parabolic models, in which the exceptional points are driven through at finite speeds which are quadratic or cubic functions of time has been analyzed by the authors \cite{kam2021non}. Compared to the two-state scenario, the research of the non-Hermitian generalization of Landau-Zener non-adiabatic transitions in the multi-state scenario is still at its early stage. Recently, the three-state non-Hermitian Landau-Zener model in the presence of an interaction with the outer environment has been considered \cite{militello2019three}, and the Landau-Zener transitions through a pair of higher order exceptional points has been analyzed by Melanathuru \cite{melanathuru2022landau}. In this work, based on our analytical approximation methods used in previous researches \cite{kam2020analytical, kam2021analytical, kam2021non}, we will analyze the dynamics of a four-state non-Hermitian $\mathcal{P}\mathcal{T}$-symmetric system which directly passes through a collection of exceptional points. 

\section{The model}
To begin with, we consider a four-state non-Hermitian system which has been used to describe the dynamics of four coupled cavities with asymmetric losses \cite{ding2016emergence}. The non-Hermitian system consists of two pairs of coupled cavities, each of which is described by a $2\times 2$ non-Hermitian Hamiltonian 
\begin{equation}\label{TwoHamiltoian}
    H_i = \left(
    \begin{array}{cccc}
   \omega_i-i\Gamma_0 & \kappa  \\
   \kappa & \omega_i-i\Gamma
  \end{array}
  \right),
\end{equation}
where $\kappa$ represents the coupling strength between the two cavities, $\omega_i$ ($i=1,2$) represents the same resonant frequency of the two cavities, and $\Gamma_0$ and $\Gamma$ represent the intrinsic loss of the each cavity. The whole non-Hermitian system consists of two pairs of above-mentioned coupled cavities with the same values of $\kappa$, $\Gamma_0$ and $\Gamma$ but different resonant frequencies $\omega_1$ and $\omega_2$. The two pairs of cavities are coupled by connecting each individual cavity of one pair with that of another pair by a small tube by an inter-pair coupling strength $\eta$. The $4\times 4$ non-Hermitian Hamiltonian of the whole system becomes
\begin{equation}\label{FourHamiltoian}
    H = \left(
    \begin{array}{cccc}
   \omega_2-i\Gamma_0 & \kappa & 0 & \eta \\
   \kappa & \omega_2-i\Gamma & \eta & 0 \\
   0 & \eta & \omega_1-i\Gamma_0 & \kappa \\
   \eta & 0 & \kappa & \omega_1-i\Gamma
  \end{array}
  \right),
\end{equation}
Evidently, the $2\times 2$ Hamiltonian for each pair of coupled cavities is $\mathcal{PT}$-symmetric only when the intrinsic losses are symmetric, i.e., $\Gamma_0=\Gamma$, where $\mathcal{P}\equiv \bigl( \begin{smallmatrix}0 & 1\\ 1 & 0\end{smallmatrix}\bigr)$ denotes the parity operator, and $\mathcal{T}$ denotes complex conjugation. From the Hamiltonian Eq.\:\eqref{FourHamiltoian}, one obtains the coupled equations for the four wave amplitudes $a_1$, $a_2$, $a_3$ and $a_4$
\begin{subequations}
\begin{align}
    i\dot{a}_1 &= (\omega_2-i\Gamma_0)a_1 +\kappa a_2 +\eta a_4,\\
    i\dot{a}_2 &= \kappa a_1+(\omega_2-i\Gamma)a_2+\eta a_3,\\
    i\dot{a}_3&=\eta a_2+(\omega_1-i\Gamma_0)a_3+\kappa a_4,\\
    i\dot{a}_4&=\eta a_1+\kappa a_3+(\omega_1-i\Gamma)a_4.
\end{align}
\end{subequations}
Using the change of variable $b_i\equiv e^{\int_0^t(\bar{\Gamma}+i\bar{\omega})ds}a_i$ to remove the average resonant frequency $\bar{\omega}\equiv \frac{1}{2}(\omega_1+\omega_2)$ and the average intrinsic loss $\bar{\Gamma}\equiv \frac{1}{2}(\Gamma+\Gamma_0)$, one obtains the new coupled equations for the wave amplitudes $b_i$, which depend only on the relative resonant frequency $\Delta\omega\equiv\omega_1-\omega_2$ and the relative intrinsic loss $\Delta\Gamma\equiv \Gamma-\Gamma_0$, as
\begin{subequations}
\begin{align}
    i\dot{b}_1 &= -\Omega^*b_1 +\kappa b_2 +\eta b_4,\\
    i\dot{b}_2 &= \kappa b_1-\Omega b_2+\eta b_3,\\
    i\dot{b}_3&=\eta b_2+\Omega b_3+\kappa b_4,\\
    i\dot{b}_4&=\eta b_1+\kappa b_3+\Omega^*b_4,
\end{align}
\end{subequations}
where $\Omega\equiv \frac{1}{2}(\Delta\omega+i\Delta\Gamma)$. One may further define $c_1\equiv b_1+ib_2$, $c_2\equiv b_1-ib_2$, $c_3\equiv b_3+ib_4$ and $c_4\equiv b_3-ib_4$, and obtains
\begin{subequations}
\begin{align}
    i\dot{c}_1&=-\Delta\omega c_1+i\gamma c_2+i\eta c_4,\label{c1}\\\label{c2}
    i\dot{c}_2&=i\gamma^\prime c_1 -\Delta\omega c_2-i\eta c_3,\\\label{c3}
    i\dot{c}_3&=i\eta c_2+\Delta\omega c_3+i\gamma c_4,\\
    i\dot{c}_4&=-i\eta c_1+i\gamma^\prime c_3+\Delta\omega c_4,\label{c4}
\end{align}
\end{subequations}
where $\gamma\equiv \Delta\Gamma+\kappa$ and $\gamma^\prime\equiv \Delta\Gamma-\kappa$. The Hamiltonian in the diabatic basis then reads
\begin{equation}
    H^\prime = \left(
    \begin{array}{cccc}
   -\Delta\omega & i\gamma & 0 & i\eta \\
   i\gamma^\prime & -\Delta\omega & -i\eta & 0 \\
   0 & i\eta & \Delta\omega & i\gamma \\
   -i\eta & 0 & i\gamma^\prime & \Delta\omega
  \end{array}
  \right).
\end{equation}
From Eqs.\:\eqref{c1} and \eqref{c2}, one immediately obtains
\begin{subequations}
\begin{align}
    c_3&=-\frac{1}{\eta}\left(\dot{c}_2-i\Delta\omega c_2-\gamma^\prime c_1\right),\label{R1}\\
    c_4&=\frac{1}{\eta}\left(\dot{c}_1-i\Delta\omega c_1-\gamma c_2\right)\label{R2}.
\end{align}
\end{subequations}
Substitution of Eqs.\:\eqref{R1} and \eqref{R2} into Eqs.\:\eqref{c3} and \eqref{c4} yields the following coupled equations
\begin{subequations}
\begin{align}
    \ddot{c}_1+\left(\eta^2+\Delta\omega^2-\gamma^{\prime 2}-i\Delta\dot{\omega}\right)c_1&=2\kappa\dot{c}_2+2i\Delta\omega\Delta\Gamma c_2,\\
    \ddot{c}_2+\left(\eta^2+\Delta\omega^2-\gamma^2-i\Delta\dot{\omega}\right)c_2&=-2\kappa\dot{c}_1+2i\Delta\omega\Delta\Gamma c_1.
\end{align}
\end{subequations}
In particular, for the $\mathcal{PT}$-symmetric case, i.e., $\Delta\Gamma=0$, one immediately obtains
\begin{subequations}
\begin{align}\label{CoupledHeun1}
    \ddot{c}_1+Q(t)c_1&=2\kappa\dot{c}_2,\\
    \ddot{c}_2+Q(t)c_2&=-2\kappa\dot{c}_1,\label{CoupledHeun2}
\end{align}
\end{subequations}
where $Q(t)\equiv\eta^2-\kappa^2+ \Delta\omega^2(t)-i\Delta\dot{\omega}(t)$. A direct computation yields
\begin{equation}
    \dot{c}_1c_2-\dot{c}_2c_1-\kappa(c_1^2+c_2^2)=\mbox{Const}.
\end{equation}
In particular, for a parabolic separation of diabatic energies, i.e., $\Delta\omega=\alpha +\beta t^2$, one will obtain
\begin{align}
    Q(t)&=\eta^2-\kappa^2+(\alpha+\beta t^2)^2-2i\beta t\nonumber\\
    &=\beta^2t^4+2\alpha \beta t^2-2i\beta t +\alpha^2+\eta^2-\kappa^2.
\end{align}
One can see that as $Q(t)$ is now a quartic function of time, in the absence of the coupling $\kappa$ within each pair of cavities, Eqs.\:\eqref{CoupledHeun1} and \eqref{CoupledHeun2} are decoupled and solved by the superposition of the tri-confluent Heun functions $T_1$ and $T_2$. When the coupling $\kappa$ is nonzero, one can write $c_1=\sum_{n=0}^\infty \kappa^n c_1^{(n)}$ and $c_2=\sum_{n=0}^\infty \kappa^n c_2^{(n)}$, where $c_1^{(0)}\equiv d_1T_1+d_2T_2$ and $c_2^{(0)}\equiv e_1T_1+e_2T_2$, and obtains the recursive relations
\begin{subequations}
\begin{align}\label{Perturbation1}
    \ddot{c}_1^{(n)}+Q(t)c_1^{(n)}&=2\dot{c}_2^{(n-1)},\\
    \ddot{c}_2^{(n)}+Q(t)c_2^{(n)}&=-2\dot{c}_1^{(n-1)}.\label{Perturbation2}
\end{align}
\end{subequations}

\section{Integrals involving products of Heun functions and their derivatives}
To solve the recursive relations, one regards the right hand side of Eqs.\:\eqref{Perturbation1} and \eqref{Perturbation2} as known functions of time, then the $n$-th order terms $c_1^{(n)}$ and $c_2^{(n)}$ are integrals of the ($n-1$)-th order terms $\dot{c}_1^{(n-1)}$ and $\dot{c}_2^{(n-1)}$, which may be expressed as
\begin{subequations}
\begin{align}\label{GeneralFormula1}
    c_1^{(n)}&= \frac{-2 T_1}{W}\int T_2\dot{c}_2^{(n-1)}dt+\frac{2 T_2}{W}\int T_1\dot{c}_2^{(n-1)}dt,\\
    c_2^{(n)}&=\frac{2 T_1}{W}\int T_2\dot{c}_1^{(n-1)}dt-\frac{2 T_2}{W}\int T_1\dot{c}_1^{(n-1)}dt.\label{GeneralFormula2}
\end{align}
\end{subequations}
where $W\equiv T_1\dot{T}_2-T_2\dot{T_1}$ is the Wronskian for $T_1$ and $T_2$, and is a constant of time. Using integration by parts, Eqs.\:\eqref{GeneralFormula1} - \eqref{GeneralFormula2} may be simplified as
\begin{subequations}
\begin{align}\label{GeneralFormula3}
    c_1^{(n)}&= \frac{2 T_1}{W}\int \dot{T}_2c_2^{(n-1)}dt-\frac{2 T_2}{W}\int \dot{T}_1c_2^{(n-1)}dt,\\
    c_2^{(n)}&=-\frac{2 T_1}{W}\int \dot{T}_2c_1^{(n-1)}dt+\frac{2 T_2}{W}\int \dot{T}_1c_1^{(n-1)}dt.\label{GeneralFormula4}
\end{align}
\end{subequations}
To continue, one needs to evaluate integrals of the products of Heun functions and their derivatives. To be more precise, let us consider the following three kinds of indefinite integrals: $\int t^n y^2dt$, $\int t^n \dot{y}ydt$ and $t^n \dot{y}^2dt$, with $y$ being any linear combination of the tri-confluent Heun functions $T_1$ and $T_2$, which satisfies the tri-confluent Heun equation $\ddot{y}+Q(t)y=0$, where $Q(t)\equiv\sum_{k=0}^4A_kt^k$ is a quartic function of time. Using the following relations $\int t^n (y_1\dot{y}_2+y_2\dot{y}_1)dt= t^n y_1y_2-n\int t^{n-1}y_1y_2dt$ and $\int t^n(y_1\dot{y}_2-y_2\dot{y}_1)dt=\frac{W}{n+1}t^{n+1}$, one immediately obtains
\begin{equation}
    \int t^n y_1\dot{y}_2dt = \frac{1}{2}\left(t^ny_1y_2+\frac{Wt^{n+1}}{n+1}-n\int t^{n-1}y_1y_2dt\right),
\end{equation}
where $y_1$ and $y_2$ are two independent solutions of the tri-confluent equation, and $W\equiv y_1\dot{y}_2-y_2\dot{y}_1$ is the Wronskian of them. The other two kinds of integrals will be more involved to evaluate. Let us define
\begin{subequations}
\begin{align}\label{HeunIntegral}
    \int t^n y_1y_2dt &\equiv P_n y_1y_2+\frac{Q_n}{2}(y_1\dot{y}_2+y_2\dot{y}_1)+R_n\dot{y}_1\dot{y}_2,\\
    \int t^n \dot{y}_1\dot{y}_2dt &\equiv L_n y_1y_2+\frac{M_n}{2}(y_1\dot{y}_2+y_2\dot{y}_1)+N_n\dot{y}_1\dot{y}_2.
\end{align}
\end{subequations}
The coefficients $P_n$, $Q_n$ and $R_n$ may be determined by taking derivative of Eq.\:\eqref{HeunIntegral}, which yields
\begin{align}
    t^ny_1y_2 &= (\dot{P}_n-QQ_n)y_1y_2+\frac{1}{2}(\dot{Q}_n+2P_n-2QR_n)(y_1\dot{y}_2+y_2\dot{y}_1)\nonumber\\
    &+(\dot{R}_n+Q_n)\dot{y}_1\dot{y}_2.
\end{align}
Hence, we obtain $\dot{P}_n-QQ_n=t^n$, $\dot{Q}_n+2P_n-2QR_n=0$ and $\dot{R}_n+Q_n=0$, which are solved by $Q_n=-\dot{R}_n$ and $P_n=\frac{1}{2}\ddot{R}_n+QR_n$, where $R_n$ satisfies the third order differential equation $\dddot{R}_n+4Q\dot{R}_n+2\dot{Q}R_n=2t^n$. To evaluate the coefficients $L_n$, $M_n$ and $N_n$, one may use the following identity
\begin{align}\label{HeunIntegral2}
    \int t^n\dot{y}_1\dot{y}_2dt &= \frac{1}{2}t^n(y_1\dot{y}_2+y_2\dot{y}_1)-\frac{n}{2}t^{n-1}y_1y_2\nonumber\\
    &+\int t^nQy_1y_2dt+\frac{n(n-1)}{2}\int t^{n-2}y_1y_2dt.
\end{align}
Substitution of Eq.\:\eqref{HeunIntegral} and $Q(t)\equiv \sum_{k=0}^4A_kt^k$ into Eq.\:\eqref{HeunIntegral2} yields
\begin{subequations}
\begin{align}
    L_n&= \sum_{k=0}^4A_kP_{n+k}+\frac{n(n-1)}{2}P_{n-2}-\frac{n}{2}t^{n-1},\\
    M_n&=\sum_{k=0}^4A_kQ_{n+k}+\frac{n(n-1)}{2}Q_{n-2}+t^n,\\
    N_n&=\sum_{k=0}^4A_kR_{n+k}+\frac{n(n-1)}{2}R_{n-2}.
\end{align}
\end{subequations}

\begin{acknowledgements}
This study was supported by the National Natural Science Foundation of China (Grant nos. 12104524).
\end{acknowledgements}

\begin{appendix}

\section{Formal solutions of $c_1^{(n)}$ and $c_2^{(n)}$}
From Eqs.\:\eqref{GeneralFormula3} and \eqref{GeneralFormula4}, a direct computation will yield
\begin{equation}
    c_1^{(1)}= t c_2^{(0)}, c_2^{(1)}=- t c_1^{(0)}.
\end{equation}
To proceed further, one can use the following integral identity with respect to any functions $y_1$ and $y_2$ and their derivatives
\begin{equation}
\int t y_1\dot{y}_2dt = \frac{t}{2}y_1y_2+\frac{Wt^2}{4}-\frac{1}{2}\int y_1y_2 dt,
\end{equation}
where $W\equiv y_1\dot{y}_2-\dot{y}_2y_1$ is the Wronskian with respect to $y_1$ and $y_2$. From this, a direct computation yields
\begin{subequations}
\begin{align}
c_1^{(2)}&=\frac{1}{2}(Q_0-t^2)c_1^{(0)}+R_0\dot{c}_1^{(0)},\\
c_2^{(2)}&= \frac{1}{2}(Q_0-t^2)c_2^{(0)}+R_0\dot{c}_2^{(0)},
\end{align}
\end{subequations}
where $Q_0\equiv -\dot{R}_0$ and $R_0$ is the solution of the third order differential equation
\begin{equation}
\frac{d^3R_0(t) }{dt^3}+ 4Q(t)\frac{dR_0(t)}{dt}+2\frac{dQ(t)}{dt}R_0(t)=2.
\end{equation}
In order to solve $c_1^{(n)}$ and $c_2^{(n)}$, one needs the following integrals
\begin{subequations}
\begin{align}
    \mathcal{L}_nT_k&=\frac{W}{2}\left[\left(\frac{t^{n+1}}{n+1}-\frac{n}{2}Q_{n-1}\right)T_k-nR_{n-1}\dot{T}_k\right]\nonumber\\
    &\equiv\frac{W}{2}(\mathcal{E}_nT_k+\mathcal{F}_n\dot{T}_k),\\
    \mathcal{L}_n\dot{T}_k&=\frac{W}{2}(M_nT_k+2N_n\dot{T}_k)=\frac{W}{2}(\mathcal{G}_nT_k+\mathcal{H}_n\dot{T}_k),
\end{align}
\end{subequations}
where $k=1,2$, and $\mathcal{L}_n\bullet\equiv T_1\int t^n \dot{T}_2 \bullet dt - T_2\int t^n\dot{T}_1 \bullet dt$. Form this, one may expand the functions $\mathcal{E}_n$, $\mathcal{F}_n$, $\mathcal{G}_n$, and $\mathcal{H}_n$ (and thus $\mathcal{L}_nT_k$ and $\mathcal{L}_n\dot{T}_k$) in series of time as
\begin{subequations}
\begin{align}\label{operator1}
    \mathcal{L}_nT_k&\equiv\frac{W}{2}\left(\sum_{l=0}^\infty E_{nl}t^lT_k+\sum_{l=0}^\infty F_{nl}t^l\dot{T}_k\right),\\
    \mathcal{L}_n\dot{T}_k&=\frac{W}{2} \left(\sum_{l=0}^\infty G_{nl}t^lT_k+\sum_{l=0}^\infty H_{nl}t^l\dot{T}_k\right).\label{operator2}
\end{align}
\end{subequations}
In general, one can express the $n$-th order correction terms $c_1^{(n)}$ and $c_2^{(n)}$ in terms of $c_1^{(0)}$, $c_2^{(0)}$, $\dot{c}_1^{(0)}$ and $\dot{c}_2^{(0)}$, and expand the coefficients in series of time as
\begin{subequations}
\begin{gather}\label{Expandsion1}
c_1^{(n)}\equiv\sum_{k=0}^\infty\left(\alpha_k^{(n)}c_1^{(0)}+\beta_k^{(n)}c_2^{(0)}+\gamma_k^{(n)}\dot{c}_1^{(0)}+\delta_k^{(n)}\dot{c}_2^{(0)}\right)t^k,\\
c_2^{(n)}\equiv \sum_{k=0}^\infty\left(\lambda_k^{(n)}c_1^{(0)}+\mu_k^{(n)}c_2^{(0)}+\nu_k^{(n)}\dot{c}_1^{(0)}+\xi_k^{(n)}\dot{c}_2^{(0)}\right)t^k.
\end{gather}
\end{subequations}
Using Eqs.\:\eqref{operator1} - \eqref{operator2}, one obtains the $(n+1)$-th order correction terms
\begin{subequations}
\begin{align}\label{Summation}
    c_1^{(n+1)}&=\frac{2}{W} \sum_{k=0}^\infty\left(\lambda_k^{(n)}\mathcal{L}_kc_1^{(0)}+\mu_k^{(n)}\mathcal{L}_kc_2^{(0)}+\nu_k^{(n)}\mathcal{L}_k\dot{c}_1^{(0)}+\xi_k^{(n)}\mathcal{L}_k\dot{c}_2^{(0)}\right)\nonumber\\
    =&\sum_{k,l=0}^\infty\left[(\lambda_k^{(n)}E_{kl}+\nu_k^{(n)}G_{kl})c_1^{(0)}+(\mu_k^{(n)}E_{kl}+\xi_k^{(n)}G_{kl})c_2^{(0)}\right.\nonumber\\
    +&\left.(\lambda_k^{(n)}F_{kl}+\nu_k^{(n)}H_{kl})\dot{c}_1^{(0)}+(\mu_k^{(n)}F_{kl}+\xi_k^{(n)}H_{kl})\dot{c}_2^{(0)}\right]t^l.\\
    c_2^{(n+1)}&=\frac{-2}{W} \sum_{k=0}^\infty\left[\alpha_k^{(n)}\mathcal{L}_kc_1^{(0)}+\beta_k^{(n)}\mathcal{L}_kc_2^{(0)}+\gamma_k^{(n)}\mathcal{L}_k\dot{c}_1^{(0)}+\delta_k^{(n)}\mathcal{L}_k\dot{c}_2^{(0)}\right]\nonumber\\
    =&-\sum_{k,l=0}^\infty\left[(\alpha_k^{(n)}E_{kl}+\gamma_k^{(n)}G_{kl})c_1^{(0)}+(\beta_k^{(n)}E_{kl}+\delta_k^{(n)}G_{kl})c_2^{(0)}\right.\nonumber\\
    +&\left.(\alpha_k^{(n)}F_{kl}+\gamma_k^{(n)}H_{kl})\dot{c}_1^{(0)}+(\beta_k^{(n)}F_{kl}+\delta_k^{(n)}H_{kl})\dot{c}_2^{(0)}\right]t^l.
\end{align}
\end{subequations}
A direct comparison between Eqs.\:\eqref{Expandsion1} and \eqref{Summation} yields the recursive relations
\begin{subequations}
\begin{align}
    \alpha_k^{(n+1)}&=\sum_{j=0}^\infty(\lambda_j^{(n)}E_{jk}+\nu_j^{(n)}G_{jk}),\\
    \beta_k^{(n+1)}&=\sum_{j=0}^\infty(\mu_j^{(n)}E_{jk}+\xi_j^{(n)}G_{jk}),\\
    \gamma_k^{(n+1)}&=\sum_{j=0}^\infty (\lambda_j^{(n)}F_{jk}+\nu_j^{(n)}H_{jk}),\\
    \delta_k^{(n+1)}&=\sum_{j=0}^\infty(\mu_j^{(n)}F_{jk}+\xi_j^{(n)}H_{jk}),\\
   \lambda_k^{(n+1)}&=-\sum_{j=0}^\infty(\alpha_j^{(n)}E_{jk}+\gamma_j^{(n)}G_{jk}),\\
    \mu_k^{(n+1)}&=-\sum_{j=0}^\infty(\beta_j^{(n)}E_{jk}+\delta_j^{(n)}G_{jk}),\\
    \nu_k^{(n+1)}&=-\sum_{j=0}^\infty (\alpha_j^{(n)}F_{jk}+\gamma_j^{(n)}H_{jk}),\\
    \xi_k^{(n+1)}&=-\sum_{j=0}^\infty(\beta_j^{(n)}F_{jk}+\delta_j^{(n)}H_{jk}).
\end{align}
\end{subequations}

\section{Recursive relation for $R_n$}
The coefficients $R_n$ for different $n$ are not independent. From the identity $\frac{d}{dt}\left(\dot{y}_1\dot{y}_2+Qy_1y_2\right)=\dot{Q}y_1y_2$, we obtain the indefinite integral $\int \dot{Q}y_1y_2dt = \dot{y}_1\dot{y}_2+Qy_1y_2$, which implies that $\sum_{k=0}^4kA_kR_{k-1}=1$. Similarly, from the identity $\frac{d}{dt}[t(\dot{y}_1\dot{y}_2+Qy_1y_2)-\frac{1}{2}(y_1\dot{y}_2+y_2\dot{y}_1)]=(2Q+t\dot{Q})y_1y_2$, we obtain the indefinite integral $\int(2Q+t\dot{Q})y_1y_2dt=t(\dot{y}_1\dot{y}_2+Qy_1y_2)-\frac{1}{2}(y_1\dot{y}_2+y_2\dot{y}_1)$, which yields $\sum_{k=0}^4(k+2)A_kR_k=t$. In general, one has the following identity $\int y_1y_2d(fQ) =f(\dot{y}_1\dot{y}_2+Qy_1y_2)-\int\dot{f}\dot{y}_1\dot{y}_2dt$ for an arbitrary function $f$. In particular, for $f=t^n$, one obtains the following indefinite integral
\begin{equation}\label{HeunIdentity}
    \int(t^n\dot{Q}+nt^{n-1}Q)y_1y_2dt = t^n(\dot{y}_1\dot{y}_2+Qy_1y_2)-n\int t^{n-1}\dot{y}_1\dot{y}_2dt.
\end{equation}
Substitution of Eq.\:\eqref{HeunIntegral2} into Eq.\:\eqref{HeunIdentity} yields
\begin{gather}
    \int (t^n\dot{Q}+2nt^{n-1}Q)y_1y_2dt = t^n(\dot{y}_1\dot{y}_2+Qy_1y_2)-\frac{n}{2}t^{n-1}(y_1\dot{y}_2+y_2\dot{y}_1)\nonumber\\\label{23}
    + \frac{n(n-1)}{2}t^{n-2}y_1y_2-\frac{n(n-1)(n-2)}{2}\int t^{n-3}y_1y_2dt.
\end{gather}
From Eq.\:\eqref{23} and using $t^n\dot{Q}+2nt^{n-1}Q=\sum_{k=0}^4(2n+k)A_kt^{n+k-1}$, one obtains the recursive relations for $R_n$ with $n\geq 3$
\begin{equation}\label{RecursiveRelationsHeun}
    \sum_{k=0}^4(2n+k)A_kR_{n+k-1}=t^n-\frac{n(n-1)(n-2)}{2}R_{n-3}.
\end{equation}
For example, for the Airy functions which satisfy $\ddot{y}+Q(t)y=0$ with $Q(t)=-t$, one recovers the recursive relation obtained by the authors previously \cite{kam2020analytical}
\begin{equation}
R_n= \frac{1}{2(2n+1)}\left[n(n-1)(n-2)R_{n-3}-2t^n\right];
\end{equation}
while for the Bessel functions which satisfy $\ddot{y}+\lambda^2 t^4y=0$ with $Q(t)=\lambda^2t^4$, one recovers the simple recursive relations \cite{kam2020analytical}
\begin{equation}
R_{n+3} = \frac{-1}{4(n+2)\lambda^2}\left[n(n-1)(n-2)R_{n-3}-2t^n\right].
\end{equation}
In general, when $A_4\neq 0$, one can obtain an explicit expression for $R_n$ in terms of $R_0$, $R_1$ and $R_2$ by revising Eq.\:\eqref{RecursiveRelationsHeun} as
\begin{align}\label{ReviseRecursive}
    R_{n+3}+\sum_{k=-2}^2g_n^{k+2}R_{n+k}=J_n,
\end{align}
where $J_n\equiv \frac{2t^n-n(n-1)(n-2)R_{n-3}}{2(2n+4)A_4}$, $g_n^{k+1}\equiv \frac{(2n+k)}{(2n+4)}\frac{A_{k}}{A_4}$, and $g_n^0\equiv 0$. Clearly, all $R_n$ may be expressed in terms of $R_0$, $R_1$ and $R_2$. Explicitly, the first few terms may be expressed as 
\begin{subequations}
\begin{align}
R_3&=J_0-\sum_{k=0}^2g_0^{k+2}R_k,\\
R_4&=J_1-g_1^4J_0-\sum_{k=0}^2h_1^{k+1}R_k,\\
R_5&=J_2-g_2^4J_1-h_2^3J_0-\sum_{k=0}^2w_2^kR_k,\\
R_6&=J_3-g_3^4J_2-h_3^3J_1-w_3^2J_0\nonumber\\
&-\sum_{k=1}^2u_3^{k-1}R_k+(g_1^1h_3^3+g_0^2w_3^2)R_0,
\end{align}
\end{subequations}
where $h_n^k\equiv g_n^k-g_n^4g_{n-1}^{k+1}$, $w_n^k\equiv h_n^k-h_n^3g_{n-2}^{k+2}$, and $u_n^k\equiv w_n^k-w_n^2g_{n-3}^{k+3}$. A direct computation yields
\begin{subequations}
\begin{align}
    R_3&=\frac{1}{4A_4}\left(1-3A_3R_2-2A_2R_1-A_1R_0\right),\\
    R_4&=\frac{t}{6A_4}-\frac{5A_3}{24A_4^2}-\left(\frac{2A_2}{3A_4}-\frac{5A_3^2}{8A_4^2}\right)R_2\nonumber\\
    &-\left(\frac{A_1}{2A_4}-\frac{5A_2A_3}{12A_4^2}\right)R_1-\left(\frac{A_0}{3A_4}-\frac{5A_1A_3}{24A_4^2}\right)R_0,\\
    R_5&=\frac{t^2}{8A_4}-\frac{7A_3t}{48A_4^2}-\frac{3A_2}{16A_4^2}+\frac{35A_3^2}{192A_4^3}\nonumber\\
    &-\left(\frac{5A_1}{8A_4}-\frac{55A_2A_3}{48A_4^2}+\frac{105A_3^3}{192A_4^3}\right)R_2\nonumber\\
    &-\left(\frac{A_0}{2A_4}-\frac{21A_1A_3}{48A_4^2}-\frac{3A_2^2}{8A_4^2}+\frac{35A_2A_3^2}{96A_4^3}\right)R_1\nonumber\\
    &+\left(\frac{7A_0A_3}{24A_4^2}+\frac{3A_1A_2}{16A_4^2}-\frac{35A_1A_3^2}{192A_4^3}\right)R_0,\\
    R_6&=\frac{t^3}{10A_4}-\frac{9A_3t^2}{80A_4^2}-\left(\frac{2A_2}{15A_4^2}-\frac{63A_3^2}{480A_4^3}\right)t\nonumber\\
    &-\frac{7A_1}{40A_4^2}+\frac{161A_2A_3}{160A_4^3}-\frac{21A_3^3}{128A_4^4}\nonumber\\
    &-\left(\frac{3A_0}{5A_4}-\frac{87A_1A_3}{80A_4^2}-\frac{8A_2^2}{15A_4^2}+\frac{49A_2A_3^2}{32A_4^3}-\frac{189A_3^4}{384A_4^4}\right)R_2\nonumber\\
    &+\left(\frac{9A_0A_3}{20A_4^2}+\frac{3A_1A_2}{4A_4^2}-\frac{63A_1A_3^2}{160A_4^3}+\frac{A_2^2A_3}{240A_4^3}-\frac{21A_2A_3^3}{64A_4^4}\right)R_1\nonumber\\
    &-\left(\frac{3}{10A_4}-\frac{4A_0A_2}{15A_4^2}-\frac{7A_1^2}{40A_4^2}+\frac{21A_0A_3^2}{80A_4^3}\right.\nonumber\\
    &\left.+\frac{161A_1A_2A_3}{480A_4^3}-\frac{315A_1A_3^3}{1920A_4^4}\right)R_0.
\end{align}
\end{subequations}

\section{Explicit expressions for $c_1^{(n)}(t)$ and $c_2^{(n)}(t)$ for the cases when $|t|\ll1$ and $|t|\rightarrow \infty$}
\subsection{Cases for $|t|\ll1$ and $Q(t)\approx -2i\beta t +\alpha^2+\eta^2-\kappa^2$}
When $|t|\ll1$, one can simply retain the linear teams in $Q(t)$ and obtains $Q(t)\approx A_1t+A_0=-2i\beta t +\alpha^2+\eta^2-\kappa^2$. After the coordinate transformation $z\equiv g(t +A_0/A_1)$, $c_1^{(n)}$ and $c_2^{(n)}$ are determined by the recursive relations
\begin{subequations}
\begin{align}
&\frac{d^2c_1^{(n)}}{dz^2}-zc_1^{(n)}=2g\frac{dc_2^{(n-1)}}{dz},\\
&\frac{d^2c_2^{(n)}}{dz^2}-zc_2^{(n)}=-2g\frac{dc_1^{(n-1)}}{dz},
\end{align}
\end{subequations}
where $g\equiv e^{i\pi/3}A_1^{1/3}$. Here, the lowest-order terms $c_1^{(0)}$ and $c_2^{(0)}$ are solved by the linear combinations of the Airy functions $\Ai(z)$ and $\Bi(z)$, i.e., $c_1^{(0)}(z)=d_1\Ai(z)+d_2\Bi(z)$ and $c_2^{(0)}(z)=e_1\Ai(z)+e_2\Bi(z)$. To proceed further, notice that the integral of the product of any linear combinations of the Airy functions has the form
\begin{equation}
\int z^n y_1y_2 dz \equiv P_ny_1y_2+\frac{Q_n}{2}(y_1y_2^\prime+y_2y_1^\prime)+R_ny_1^\prime y_2^\prime,
\end{equation}
where $P_n=\frac{1}{2}R_n^{\prime\prime}-zR_n$, $Q_n=-R_n^\prime$ and $R_n$ is determined by the third-order differential equation
\begin{equation}
\frac{d^3R_n}{dz^3}-4z\frac{dR_n}{dz}-2R_n=2z^n.
\end{equation}
A straightforward computation yields $R_n$, $Q_n$ and $P_n$ for $n\leq 2$
\begin{subequations}
\begin{gather}
R_0=-1, Q_0=0, P_0=z,\\
R_1=-\frac{z}{3},Q_1=\frac{1}{3},P_1=\frac{z^2}{3},\\
R_2=-\frac{z^2}{5},Q_2=\frac{2z}{5},P_2=\frac{z^3-1}{5},
\end{gather}
\end{subequations}
where $R_n$ for $n\geq 3$ is determined by the recursive relation
\begin{equation}
R_n = -\frac{z^n}{2n+1} +\frac{n(n-1)(n-2)}{2(2n+1)}R_{n-3}.
\end{equation}
Hence, $R_n$ for $n\geq 3$ is solved by
\begin{align}
R_n &= -\frac{z^n}{2n+1} -\frac{n(n-1)(n-2)z^{n-3}}{2(2n+1)(2n-5)} \nonumber\\
&-\cdots -\frac{n(n-1)\cdots(n-3k+1)z^{n-3k}}{2^k(2n+1)\cdots(2n+1-6k)}\nonumber\\
&=-\frac{z^n}{2n+1}\sum_{j=0}^k \frac{n!\Gamma(\frac{2n+1}{6})(12z^3)^{-j}}{(n-3j)!\Gamma(\frac{2n+1}{6}-j)},
\end{align}
where $k$ is the number such that $n-3k\in [0,2]$. Using the relation $Q_n=-\dot{R}_n$, a direct computation yields
\begin{align}
Q_n &= \frac{nz^{n-1}}{2n+1} +\frac{n(n-1)(n-2)(n-3)z^{n-4}}{2(2n+1)(2n-5)} \nonumber\\
&+\cdots +\frac{n(n-1)\cdots(n-3l)z^{n-3l-1}}{2^k(2n+1)\cdots(2n+1-6l)}\nonumber\\
&=\frac{nz^{n-1}}{2n+1}\sum_{j=0}^l \frac{(n-1)!\Gamma(\frac{2n+1}{6})(12z^3)^{-j}}{(n-1-3j)!\Gamma(\frac{2n+1}{6}-j)},
\end{align}
where $l$ is the number such that $n-3l-1\in [0,2]$. Using the above relations, $R_n$, $Q_n$ and $P_n$ for $n\leq 6$ can be explicitly expressed as
\begin{subequations}
\begin{gather}
R_3=-\frac{z^3+3}{7}, Q_3=\frac{3z^2}{7}, P_3 = \frac{z^4}{7},\\
R_4=-\frac{z^4+4z}{9},Q_4=\frac{4z^3+4}{9},P_4=\frac{z^5-2z^2}{9},\\
R_5=-\frac{z^5+6z^2}{11},Q_5=\frac{5z^4+12z}{11},P_5=\frac{z^6-4z^3-6}{11}.
\end{gather}
\end{subequations}

\subsection{Cases for $|t|\gg 1$ and $Q(t)\approx \beta^2t^4$}
In the region $|t|\gg 1$, one can only keep the highest order term in $Q(t)$ such that $Q(t)\approx \beta^2t^4$. Then, $c_1^{(n)}(t)$ and $c_2^{(n)}(t)$ are determined by the recursive equations
\begin{subequations}
\begin{align}
\ddot{c}_1^{(n)}&+\beta^2t^4c_1^{(n)}=2\dot{c}_2^{(n-1)},\\
\ddot{c}_2^{(n)}&+\beta^2t^4c_2^{(n)}=-2\dot{c}_1^{(n-1)}.
\end{align}
\end{subequations}
Here the zeroth order terms $c_1^{(0)}(t)$ and $c_2^{(0)}(t)$ are linear combination of $\bar{y}_1\equiv \sqrt{t}J_{1/6}(\beta t^3/3)$ and $\bar{y}_2\equiv \sqrt{t}J_{-1/6}(\beta t^3/3)$, where $J_\nu(z)$ are the Bessel function of the first kind defined by
\begin{equation}
J_{\nu}(z)\equiv \sum_{n=0}^\infty \frac{(-1)^n}{\Gamma(\nu+n+1)n!}\left(\frac{z}{2}\right)^{\nu+2n}.
\end{equation}
Hence, the fundamental solutions $y_1(t)$ and $y_2(t)$ of the equation $\ddot{y}+\beta^2t^4y=0$ and their derivatives have the series expansions
\begin{subequations}
\begin{align}
y_1 &\equiv\Gamma\left(\frac{7}{6}\right)\left(\frac{\beta}{6}\right)^{-\frac{1}{6}}\bar{y}_1= \sum_{n=0}^\infty \frac{\Gamma(\frac{7}{6})(-1)^n}{\Gamma(\frac{7}{6}+n)n!}\left(\frac{\beta}{6}\right)^{2n}t^{1+6n},\\
y_2&\equiv\Gamma\left(\frac{5}{6}\right)\left(\frac{\beta}{6}\right)^{\frac{1}{6}}\bar{y}_2=\sum_{n=0}^\infty \frac{\Gamma(\frac{5}{6})(-1)^n}{\Gamma(\frac{5}{6}+n)n!}\left(\frac{\beta}{6}\right)^{2n}t^{6n},\\
\dot{y}_1&=\sum_{n=0}^\infty \frac{\Gamma(\frac{1}{6})(-1)^n}{\Gamma(\frac{1}{6}+n)n!}\left(\frac{\beta}{6}\right)^{2n}t^{6n},\\
\dot{y}_2 &=-\frac{\beta^2t^5}{5}\sum_{n=0}^\infty \frac{\Gamma(\frac{11}{6})(-1)^{n+1}}{\Gamma(\frac{11}{6}+n)n!}\left(\frac{\beta}{6}\right)^{2n}t^{6n}.
\end{align}
\end{subequations}
The fundamental solutions $y_1(t)$ and $y_2(t)$ and their derivatives can be expressed in terms of the generalized hypergeometric functions ${}_pF_q(a_1,\cdots,a_p;b_1,\cdots,b_q;z)$ as
\begin{subequations}
\begin{align}
y_1(t)&={}_1F_2\left(1;1,\frac{7}{6};\frac{-\beta^2t^6}{36}\right)t,\\
y_2(t)&={}_1F_2\left(1;1,\frac{5}{6};\frac{-\beta^2t^6}{36}\right),\\
\dot{y}_1(t)&={}_1F_2\left(1;1,\frac{1}{6};\frac{-\beta^2t^6}{36}\right),\\
\dot{y}_2(t)&=-{}_1F_2\left(1;1,\frac{11}{6};\frac{-\beta^2t^6}{36}\right)\frac{\beta^2t^5}{5},
\end{align}
\end{subequations}
which obey the initial conditions $y_1(0)=0$, $y_2(0)=1$, $\dot{y}_1(0)=1$ and $\dot{y}_2(0)=0$. Hence, the Wronskian with respect to $y_1$ and $y_2$ is a constant, i.e., $W(y_1,y_2)\equiv y_1\dot{y}_2-y_2\dot{y}_1=-1$. A direct computation yields the products of the fundamental solutions and their derivatives 
\begin{align}
&y_1y_2 ={}_1F_2(1;1,\frac{5}{6};\frac{-\beta^2t^6}{36}){}_1F_2(1;1,\frac{7}{6};\frac{-\beta^2t^6}{36})t,\nonumber\\
&\dot{y}_1\dot{y}_2=-{}_1F_2(1;1,\frac{1}{6};\frac{-\beta^2t^6}{36}){}_1F_2(1;1,\frac{11}{6};\frac{-\beta^2t^6}{36})\frac{\beta^2t^5}{5},\nonumber\\
&y_1\dot{y}_2+y_2\dot{y}_1={}_1F_2(1;1,\frac{1}{6};\frac{-z^2}{4}){}_1F_2(1;1,\frac{5}{6};\frac{-z^2}{4})\nonumber\\
&-{}_1F_2(1;1,\frac{7}{6};\frac{-z^2}{4}){}_1F_2(1;1,\frac{11}{6};\frac{-z^2}{4})\frac{\beta^2t^6}{5}.
\end{align}
Thus, the above products can be expanded in the Taylor series as
\begin{align}
y_1y_2 &=t-\frac{2}{35}\beta^2t^7+\frac{6}{5005}\beta^4t^{13}+O(t^{19}),\nonumber\\
y_1\dot{y}_2+y_2\dot{y}_1&=1-\frac{2}{5}\beta^2t^6+\frac{6}{385}\beta^4t^{12}+O(t^{18}),\nonumber\\
\dot{y}_1\dot{y}_2&=-\frac{1}{5}\beta^2t^5+\frac{2}{55}\beta^4t^{11}+O(t^{17}).
\end{align}
Similar to the previous case, the integral of the product of the fundamental solutions $y_1(t)$ and $y_2(t)$ with a power weight $t^n$ can be written in the form
\begin{equation}
\int t^n y_1y_2 dt \equiv P_ny_1y_2+\frac{Q_n}{2}(y_1\dot{y}_2+y_2\dot{y}_1)+R_n\dot{y}_1\dot{y}_2,
\end{equation}
where $P_n=\frac{1}{2}\ddot{R}_n+\beta^2t^4R_n$, $Q_n=-\dot{R}_n$, and $R_n$ is determined by the third-order differential equation
\begin{equation}\label{ThirdOrderDiffEq}
\frac{d^3R_n}{dt^3}+4\beta^2t^4\frac{dR_n}{dt}+8\beta^2t^3R_n=2t^n,
\end{equation}
from which one can show that $R_n$ obey the recursive relation
\begin{equation}\label{BesselRecursive}
R_{n+6}=\frac{t^{n+3}}{2\beta^2(n+5)}-\frac{(n+3)(n+2)(n+1)}{4\beta^2(n+5)}R_{n},
\end{equation}
A direct computation gives $R_n$, $Q_n$ and $P_n$ for $n=3,4,5$
\begin{subequations}
\begin{align}
R_3(t) = \frac{1}{4\beta^2},&Q_3(t)=0,P_3(t)=\frac{t^4}{4},\\
R_4(t) = \frac{t}{6\beta^2},&Q_4(t)=-\frac{1}{6\beta^2},P_4(t)=\frac{t^5}{6},\\
R_5(t) = \frac{t^2}{8\beta^2},&Q_5(t)=-\frac{t}{4\beta^2},P_5(t)=\frac{t^6}{8}+\frac{1}{8\beta^2}.
\end{align}
\end{subequations}
Then, one only need to compute $R_n(t)$ for $n=0$, $1$ and $2$. To proceed further, one can expand the function $R_0(t)$ as $R_0(t)=\sum_{k=0}^\infty r_kt^k$. From Eq.\:\eqref{ThirdOrderDiffEq}, one immediately obtains the recursive relation
\begin{equation}
r_{k+6}=\frac{-4\beta^2(k+2)}{(k+6)(k+5)(k+4)}r_k,
\end{equation}
For $n=0$, the functions $R_0$, $Q_0$ and $P_0$ can be expressed in terms of the generalized hypergeometric function as
\begin{subequations}
\begin{align}
R_0(t)&={}_2F_3\left(1,\frac{5}{6};\frac{7}{6},\frac{8}{6},\frac{9}{6};-\frac{\beta^2t^6}{9}\right)\frac{t^3}{3},\\
Q_0(t)&=-{}_2F_3\left(1,\frac{5}{6};\frac{3}{6},\frac{7}{6},\frac{8}{6};-\frac{\beta^2t^6}{9}\right)t^2,\\
P_0(t)&={}_2F_3\left(1,\frac{5}{6};\frac{2}{6},\frac{3}{6},\frac{7}{6};-\frac{\beta^2t^6}{9}\right)t\nonumber\\
&+{}_2F_3\left(1,\frac{5}{6};\frac{7}{6},\frac{8}{6},\frac{9}{6};-\frac{\beta^2t^6}{9}\right)\frac{\beta^2t^7}{3}.
\end{align}
\end{subequations}
These functions define different entire functions of $t$, which has the following series expansions
\begin{align}
R_0(t)&= \frac{1}{3}t^3-\frac{5}{378}\beta^2t^9+\frac{11}{51597}\beta^4t^{15}+O(t^{21}),\nonumber\\
Q_0(t)&= -t^2+\frac{5}{42}\beta^2t^8-\frac{55}{17199}\beta^4t^{14}+O(t^{20}),\nonumber\\
P_0(t)&=t-\frac{1}{7}\beta^2t^7+\frac{5}{546}\beta^4t^{13}+O(t^{19}).
\end{align}
For $n=1$, the functions $R_1$, $Q_1$ and $P_1$ can be expressed in terms of the generalized hypergeometric function as
\begin{subequations}
\begin{align}
R_1(t)&={}_2F_3\left(1,1;\frac{8}{6},\frac{9}{6},\frac{10}{6};-\frac{\beta^2t^6}{9}\right)\frac{t^4}{12},\\
Q_1(t)&=-{}_2F_3\left(1,1;\frac{4}{6},\frac{8}{6},\frac{9}{6};-\frac{\beta^2t^6}{9}\right)\frac{t^3}{3},\\
P_1(t)&={}_2F_3\left(1,1;\frac{3}{6},\frac{4}{6},\frac{8}{6};-\frac{\beta^2t^6}{9}\right)\frac{t^2}{2}\nonumber\\
&+{}_2F_3\left(1,1;\frac{8}{6},\frac{9}{6},\frac{10}{6};-\frac{\beta^2t^6}{9}\right)\frac{\beta^2t^8}{12}.
\end{align}
\end{subequations}
These functions define different entire functions of $t$, which has the following series expansions
\begin{align}
R_1(t)&= \frac{1}{12}t^4-\frac{1}{360}\beta^2t^{10}+\frac{1}{25200}\beta^4t^{16}+O(t^{22}),\nonumber\\
Q_1(t)&= -\frac{t^3}{3}+\frac{1}{36}\beta^2t^9-\frac{1}{1575}\beta^4t^{15}+O(t^{21}),\nonumber\\
P_1(t)&=\frac{t^2}{2}-\frac{1}{24}\beta^2t^8+\frac{1}{504}\beta^4t^{14}+O(t^{20}).
\end{align}
For $n=2$, the functions $R_2$, $Q_2$ and $P_2$ can be expressed in terms of the generalized hypergeometric function as
\begin{subequations}
\begin{align}
R_2(t)&={}_2F_3\left(1,\frac{7}{6};\frac{9}{6},\frac{10}{6},\frac{11}{6};-\frac{\beta^2t^6}{9}\right)\frac{t^5}{30},\\
Q_2(t)&=-{}_2F_3\left(1,\frac{7}{6};\frac{5}{6},\frac{9}{6},\frac{10}{6};-\frac{\beta^2t^6}{9}\right)\frac{t^4}{6},\\
P_2(t)&={}_2F_3\left(1,\frac{7}{6};\frac{4}{6},\frac{5}{6},\frac{9}{6};-\frac{\beta^2t^6}{9}\right)\frac{t^3}{3}\nonumber\\
&+{}_2F_3\left(1,\frac{7}{6};\frac{9}{6},\frac{10}{6},\frac{11}{6};-\frac{\beta^2t^6}{9}\right)\frac{\beta^2t^9}{30}.
\end{align}
\end{subequations}
These functions define different entire functions of $t$, which has the following series expansions
\begin{align}
R_2(t)&= \frac{1}{30}t^5-\frac{7}{7425}\beta^2t^{11}+\frac{91}{7573500}\beta^4t^{17}+O(t^{23}),\nonumber\\
Q_2(t)&= -\frac{t^4}{6}+\frac{7}{675}\beta^2t^{10}-\frac{91}{445500}\beta^4t^{16}+O(t^{22}),\nonumber\\
P_2(t)&=\frac{t^3}{3}-\frac{1}{54}\beta^2t^9+\frac{7}{10125}\beta^4t^{15}+O(t^{21}).
\end{align}
From the recursive relation Eq.\:\eqref{BesselRecursive}, one can derive $R_n(t)$, $Q_n(t)$ and $P_n(t)$ for $n=6k+m$ with $m=0,1,2$, and $k$ being any non-negative integer
\begin{align}
R_{n}(t)&=\frac{2t^{n+3}{}_2F_3\left(1,\frac{n+5}{6};\frac{n+7}{6},\frac{n+8}{6},\frac{n+9}{6};-\frac{\beta^2t^6}{9}\right)}{(n+1)(n+2)(n+3)},\nonumber\\
Q_{n}(t)&=-\frac{2t^{n+2}{}_2F_3\left(1,\frac{n+5}{6};\frac{n+3}{6},\frac{n+7}{6},\frac{n+8}{6};-\frac{\beta^2t^6}{9}\right)}{(n+1)(n+2)},\nonumber\\
P_n(t)&=\frac{t^{n+1}{}_2F_3\left(1,\frac{n+5}{6};\frac{n+2}{6},\frac{n+3}{6},\frac{n+7}{6};-\frac{\beta^2t^6}{9}\right)}{n+1}\nonumber\\
&+\frac{2\beta^2t^{n+7}{}_2F_3\left(1,\frac{n+5}{6};\frac{n+7}{6},\frac{n+8}{6},\frac{n+9}{6};-\frac{\beta^2t^6}{9}\right)}{(n+1)(n+2)(n+3)}.
\end{align}
With the analytical expressions of $R_n$, $Q_n$ and $P_n$, one can systematically derive $c_1^{(n)}$ and $c_2^{(n)}$. For example, for $n=2$, one obtains
\begin{subequations}
\begin{align}
c_1^{(2)}(t)&=-\frac{t^2}{2}\left[1+{}_2F_3\left(1,\frac{5}{6};\frac{3}{6},\frac{7}{6},\frac{8}{6};-\frac{\beta^2t^6}{9}\right)\right]c_1^{(0)}(t)\nonumber\\
&+\frac{t^3}{3}{}_2F_3\left(1,\frac{5}{6};\frac{7}{6},\frac{8}{6},\frac{9}{6};-\frac{\beta^2t^6}{9}\right)\dot{c}_1^{(0)}(t),\\
c_2^{(2)}(t)&=-\frac{t^2}{2}\left[1+{}_2F_3\left(1,\frac{5}{6};\frac{3}{6},\frac{7}{6},\frac{8}{6};-\frac{\beta^2t^6}{9}\right)\right]c_2^{(0)}(t)\nonumber\\
&+\frac{t^3}{3}{}_2F_3\left(1,\frac{5}{6};\frac{7}{6},\frac{8}{6},\frac{9}{6};-\frac{\beta^2t^6}{9}\right)\dot{c}_2^{(0)}(t).
\end{align}
\end{subequations}

\end{appendix}

\end{document}